\documentclass[12pt,a4paper]{article}
\usepackage{graphicx}
\usepackage[T1]{fontenc}
\usepackage[utf8]{inputenc}
\usepackage{textcomp}
\usepackage[sc,osf]{mathpazo}
\usepackage{a4wide}  
\usepackage{latexsym,amsthm,amsfonts,amsmath,mathrsfs,amssymb}
\usepackage{dsfont}
\usepackage{accents}
\usepackage[nosort]{cite}
\usepackage{booktabs} 
\usepackage[unicode,implicit]{hyperref}
\hypersetup{%
  pdftitle    = {Gravitational higher--form symmetries and the origin
    of hidden symmetries in Kaluza--Klein compactifications}
  pdfkeywords = {~},
  pdfauthor   = {Carmen G\'omez-Fayr\'en, Tom\'as Ort\'{\i}n and Matteo Zatti},
  plainpages  = true,
  colorlinks  = true,
  citecolor   = blue,
  urlcolor    = red,
  linkcolor   = black
}
\newcommand{\hepth}[1]{{\tt
\href{http://www.arXiv.org/abs/hep-th/#1}{hep-th/#1}}}

\newcommand{\arxiv}[1]{{\tt arXiv:\href{http://www.arXiv.org/abs/#1}{#1}}}
\newcommand{\doi}[1]{{\tt DOI:\href{http://dx.doi.org/#1}{#1}}}

\allowdisplaybreaks

\makeatletter
\@addtoreset{equation}{section}
\makeatother

\begin{document}

\begin{flushright}
\small
IFT-UAM/CSIC-24-076\\
May 27\textsuperscript{th}, 2024\\
\normalsize
\end{flushright}

\vspace{1cm}

\begin{center}

  {\Large {\bf Gravitational higher--form symmetries}}\\[.5cm]
  {\Large {\bf and the origin of hidden symmetries}}\\[.5cm]
  {\Large {\bf in Kaluza--Klein compactifications}}

\vspace{2.5cm}

\renewcommand{\thefootnote}{\alph{footnote}}

{\sl\large Carmen G\'omez-Fayr\'en},\footnote{Email: {\tt carmen.gomez-fayren[at]estudiante.uam.es}}
{\sl\large Tom\'as Ort\'{\i}n}\footnote{Email: {\tt  tomas.ortin[at]csic.es}}
{\sl\large and Matteo Zatti}\footnote{Email: {\tt matteo.zatti[at]estudiante.uam.es}}

\setcounter{footnote}{0}
\renewcommand{\thefootnote}{\arabic{footnote}}
\vspace{1cm}

{\it Instituto de F\'{\i}sica Te\'orica UAM/CSIC\\[10pt]
C/ Nicol\'as Cabrera, 13--15,  C.U.~Cantoblanco, E-28049 Madrid, Spain}\\

\vspace{3cm}


{\bf Abstract}
\end{center}
\begin{quotation}
  {\small We show that, in presence of isometries and non-trivial topology,
    the Einstein--Hilbert action is invariant under certain transformations of
    the metric which are not diffeomorphisms. These transformations are
    similar to the higher-form symmetries of field theories with $p$-form
    fields. In the context of toroidal Kaluza--Klein compactifications, we
    show that these symmetries give rise to some of the ``hidden symmetries''
    (dualities) of the dimensionally-reduced theories.}
\end{quotation}

\newpage
\pagestyle{plain}




\section{Introduction}

The main theme of Kaluza--Klein (KK) theories\footnote{Many historical and
  more modern references on these theories can be found in
  Ref.~\cite{Appelquist:1987nr}.} is that the symmetries of the
lower-dimensional theory obtained by compactification have a
higher-dimensional purely gravitational (that is, geometric) origin.  It is
not difficult to see how families of diffeomorphisms of the higher-dimensional
spacetime that depend on a number of arbitrary functions give rise to the
Yang--Mills-type gauge symmetries of the lower-dimensional theory. Most of the
lower-dimensional theories obtained by KK compactification also have global
symmetries which, in the Einstein frame, only act on the matter fields. The
origin of these global symmetries, formerly known as ``hidden symmetries''
and, more recently in the context of superstring/supergravity theories, as
``dualities'', is more mysterious.\footnote{Explaining and making manifest all
  the global symmetries found in lower-dimensional string effective theories
  (supergravities) from the 10- and 11-dimensional point of view is the goal
  of double and exceptional field theory. For a recent review, see
  Ref.~\cite{Samtleben:2023nwk} and references therein.}

There are two main kinds of hidden symmetries:

\begin{enumerate}
\item Those which involve electric-magnetic-type duality transformations,
  which, typically, only leave invariant the equations of motion and can be
  considered non-perturbative.
\item The rest, which leave the action invariant.
\end{enumerate}

The higher-dimensional origin of the former is unknown\footnote{As we are
  going to explain, the standard explanation is not completely correct.} and
our results do not seem to provide new information about it. Thus, we are
going to focus on the later, and, in particular, on symmetries which do not
mix fields originating in different higher-dimensional fields.\footnote{In
  particular, this excludes T~duality
  \cite{Alvarez:1994dn,Buscher:1987sk,Buscher:1987qj,Bergshoeff:1994dg,Bergshoeff:1995as,Bergshoeff:1995cg,Meessen:1998qm,Elgood:2020xwu}.}
For the sake of clarity, in this paper we will just consider pure
$\hat{d}$-dimensional\footnote{We indicate with hats all higher-dimensional
  objects, except for the Killing vector $k$ and the vector $\epsilon$:
  $\{\hat{x}^{\hat{\mu}}\} =\{x^{\mu},z \}$ etc. We use the notation and
  conventions of Ref.~\cite{Ortin:2015hya}.} gravity, described by the
Einstein--Hilbert (EH) action, and its toroidal compactifications.

In most of the literature, the global symmetries of the theories obtained in
these toroidal compactifications are understood as simple translations,
rotations and rescalings (diffeomorphisms) in the internal directions. Closer
inspection, though, shows that many of these diffeomorphisms are incompatible
with the boundary conditions in the internal directions. Therefore, they are
not diffeomorphisms of the toroidally compactified manifolds.

Let us start by considering the compactification on a circle, from $\hat{d}$
to $d=\hat{d}-1$ dimensions of the EH action

\begin{equation}
  \label{eq:higherEHaction}
  S_{EH}[\hat{g}]
  \sim
  \int d^{\hat{d}}\hat{x}\,\sqrt{|\hat{g}|}\, \hat{R}(\hat{g})\,.
\end{equation}

\noindent
If we parametrize the compact direction with the periodic coordinate
$z\sim z+2\pi \ell$ adapted to the isometry $k=\partial_{\underline{z}}$, the
$\hat{d}$-dimensional metric decomposes into a metric $g_{\mu\nu}$, a gauge
field (the \textit{KK vector}) $A=A_{\mu}dx^{\mu}$ and a scalar (the
\textit{KK scalar}) $k$

\begin{equation}
  d\hat{s}^{2}
  =
  \hat{g}_{\hat{\mu}\hat{\nu}}d\hat{x}^{\hat{\mu}}d\hat{x}^{\hat{\nu}}
  =
 ds^{2}-k^{2}(dz+A)^{2}\,,
\end{equation}

\noindent
where

\begin{equation}
  ds^{2}
  =
  g_{\mu\nu}dx^{\mu}dx^{\nu}\,.
\end{equation}

After integration over $z$, the $\hat{d}$-dimensional EH action
can be rewritten, up to total derivatives, in the form

\begin{equation}
  S[g,A,k]
  \sim
  \int d^{d}x\,\sqrt{|g|}\,
  \left\{ kR(g) -\tfrac{1}{4}k^{3}F^{2}\right\}\,,
\end{equation}

\noindent
where

\begin{equation}
  F^{2}
  =
  F_{\mu\nu} F^{\mu\nu}\,,
  \hspace{1.5cm}
  F_{\mu\nu}
  =
  2\partial_{[\mu}A_{\nu]}\,.
\end{equation}

The factor $k$ in front of the Ricci scalar indicates that the metric
$g_{\mu\nu}$ is not the Einstein-frame metric $g_{E\, \mu\nu}$. It is,
however, related to it by

\begin{equation}
  g_{\mu\nu}
  =
  k^{-2/(\hat{d}-3)}g_{E\, \mu\nu}
  =
  k^{-2/(d-2)}g_{E\, \mu\nu}\,,
\end{equation}

\noindent
and, if we rewrite the metric in the Einstein-frame, it takes the form

\begin{equation}
  \label{eq:Einsteinframeactioncirclecompactification}
  S[g_{E},A,k]
  \sim
  \int d^{d}x\,\sqrt{|g_{E}|}\,
  \left\{ R(g_{E})+\frac{d-1}{d-2}k^{-2}(\partial k)^{2}
    -\tfrac{1}{4}k^{2\frac{d-1}{d-2}}F^{2}\right\}\,.
\end{equation}

This action is invariant under global rescalings of $k$ and $A$

\begin{equation}
  \label{eq:alphatransformationsAk}
  \delta_{\alpha}A
  =
  \alpha A\,,
  \hspace{1cm}
  \delta_{\alpha}k
  =
  -\alpha\frac{d-2}{d-1}k\,,
\end{equation}

\noindent
which do not act on the Einstein-frame metric.

As discussed, for instance, in Ref.~\cite{Ortin:2015hya}, these
transformations may be seen as the combination of a global rescaling of the
coordinate $z$ and a global rescaling of the $\hat{d}$-dimensional metric

\begin{equation}
  \label{eq:alphatransform}
  \delta_{\alpha}
  \hat{g}_{\hat{\mu}\hat{\nu}}
  =
  -2\alpha \delta^{\underline{z}}{}_{(\hat{\mu}}\hat{g}_{\hat{\nu})\underline{z}}
  +\frac{2\alpha}{\hat{d}-2} \hat{g}_{\hat{\mu}\hat{\nu}}\,,
\end{equation}

\noindent
where $\alpha$ is some constant, infinitesimal, parameter.

The rescaling of the coordinate $z$ is usually seen as a diffeomorphism
generated by the vector

\begin{equation}
  \hat{\eta}
  \equiv
  z\partial_{\underline{z}}\,,
\end{equation}

\noindent
so the first term in the transformation Eq.~(\ref{eq:alphatransform}) is just

\begin{equation}
-\alpha\pounds_{\hat{\eta}}\hat{g}_{\hat{\mu}\hat{\nu}}\,,  
\end{equation}

\noindent
where $\pounds_{\hat{\eta}}$ stands for the Lie derivative with respect to the
vector field $\hat{\eta}$.

Such a rescaling, however, changes the periodicity of $z$ when it is periodic
and cannot be consistently used in this setting. Another way of seeing this
problem is to observe that the vector field $\hat{\eta}$ is not well defined
when $z$ is periodic: it is not single valued. Furthermore, observe that,
according to this interpretation, the complete transformation
Eq.~(\ref{eq:alphatransform}) would combine a symmetry of the action up to
total derivatives (a diffeomorphism) with a transformation which is not a
symmetry of the action (a global rescaling of the metric). Such a combination
should not be a symmetry of the $\hat{d}$-dimensional theory!

As we are going to show, there is another, consistent, way of interpreting
this transformation in which the first term in Eq.~(\ref{eq:alphatransform})
corresponds to a transformation which is not a diffeomorphism nor a symmetry
of the action, because it rescales it. This rescaling can be compensated with
a global rescaling of the metric, resulting in a global symmetry of the
$\hat{d}$-dimensional action which is inherited by the $d$-dimensional one and
which corresponds, precisely, to the global symmetry in
Eq.~(\ref{eq:alphatransformationsAk}).

In toroidal compactifications, when there is more than one compact direction,
the lower-dimensional theory is usually invariant under rescalings and also
under rotations of the fields. They are usually understood as originating in
rotations of the internal manifold. However, the putative
$\hat{d}$-dimensional vector fields that would generated those rotations also
have components proportional to compact coordinates and they are not single
valued. Again, the rotations of the lower-dimensional fields are not
diffeomorphisms in higher dimensions.

As we are going to see, the transformations which is not diffeomorphisms and
which are associated to the rescalings and rotations of the compact
coordinates are the analog of the higher-form symmetries of theories of
$(p+1)$-form fields \cite{Gaiotto:2014kfa}.\footnote{For pedagogical reviews
  with many references, see, for instance
  Refs.~\cite{Gomes:2023ahz,Bhardwaj:2023kri}.} They are present in spaces of
non-trivial topology. The symmetries that we have found also require an
isometry, but they should be considered the simplest example of this kind of
symmetries and an invitation to explore other possibilities.

This paper is organized as follows: in Section~\ref{sec-higher} we review
higher-form symmetries in theories of $(p+1)$-form fields from the purely
classical point of view in order to clarify in which sense the symmetries of
the EH action that we are going to find in Section~\ref{sec-highergrav} are
similar. In Section~\ref{sec-higherKK} we show how some of these symmetries
give rise to the hidden symmetries that arise in toroidal KK
compactifications. We discuss our results in Section~\ref{sec-discussion}
indicating possible extensions and future lines of research.

\section{Higher-form symmetries}
\label{sec-higher}

In this section we are going to review the so-called higher-form symmetries of
$(p+1)$-form fields $A$ from a purely classical point of view. We will use a
language close to the one we will employ in the gravity case which is our main
interest and we will consider the simplest setting in which there is only one
field of this kind in the theory and there are no Chern--Simons-like terms
neither in the action nor in the field strength, which is, just, the
$(p+2)$-form $F=dA$.  It is convenient to use the language of differential
forms in which the manifestly gauge-invariant action in $d$ dimensions takes
the form

\begin{equation}
  S[A]
  =
  \int_{\mathcal{M}} \frac{(-1)^{(p+1)d}}{2} F\wedge \star F\,.
\end{equation}

Under an arbitrary variation of $A$

\begin{equation}
  \label{eq:genericvariationp+1formtheory}
  \delta S[A]
    =
    \int_{\mathcal{M}}\left\{\mathbf{E}_{A}\wedge \delta A
      +d\mathbf{\Theta}(A,\delta A) \right\}\,,
\end{equation}

\noindent
where

\begin{subequations}
  \begin{align}
\mathbf{E}_{A}
    & =
      - d \star F\,,
    \\
    & \nonumber \\
  \mathbf{\Theta}(A,\delta A)
    & =
     \star F \wedge \delta A\,,
  \end{align}
\end{subequations}

\noindent
are, respectively, the equation of motion of $A$ and the pre-symplectic
potential.

The field strength $F$ is invariant under the transformations $\delta A$ which
are closed

\begin{equation}
  \delta F
  =
  d\delta A
  =
  0\,.
\end{equation}

In a generic manifold $\mathcal{M}$, we can only count on the variations
$\delta$ which are exact, \textit{i.e.}

\begin{equation}
  \label{eq:exacttransformationsA}
  \delta_{\chi}A
  =
  d\chi\,,
\end{equation}

\noindent
where $\chi$ is an arbitrary $p$-form field. These are the standard gauge
transformations of a $(p+1)$-form field. However, if $\mathcal{M}$ has some
prescribed topology, there may be closed $(p+1)$-forms $\delta A$ which are
not exact. In some cases (typically, in compact manifolds $\mathcal{M}$) we
can find a basis of these forms $\{h^{(p+1)}_{i}\}$, and, then, both $F$ and
the theory will be exactly invariant under linear combinations of them with
arbitrary, constant, coefficients $a^{i}$:

\begin{equation}
  \label{eq:deltaaA}
  \delta_{a}A
  =
  a^{i}h^{(p+1)}_{i}
  \equiv
  a^{i}\delta_{i}A\,.
\end{equation}

These are global symmetries of the action and we must study them separately
from the gauge ones.

Let us consider them first. The exact invariance of the action for any
$\mathcal{M}$ in the same topology class implies the relation

\begin{equation}
  d\mathbf{\Theta}(A,\delta_{a} A)
  =
  -\mathbf{E}_{A}\wedge \delta_{a} A\,,
\end{equation}

\noindent
and for each independent global parameters $a^{i}$, we derive from it a
relation of the form

\begin{subequations}
  \begin{align}
  d \mathbf{J}_{i}
  & \doteq
  0\,,
  \\
    & \nonumber \\
    \mathbf{J}_{i}
    & =
      \mathbf{\Theta}(A,\delta_{i}A)
      =
      \star F \wedge h^{(p+1)}_{i}\,.
  \end{align}
\end{subequations}

Each $(d-1)$-form $\mathbf{J}_{i}$ is the Hodge dual of the standard Noether
current 1-form associated to the global invariance generated by the parameter
$a^{i}$. The charges associated to these currents are computed by integrating
over $(d-1)$-dimensional ``volumes''.

In the case of the gauge transformations Eq.~(\ref{eq:exacttransformationsA}),
particularizing the generic variation
Eq.~(\ref{eq:genericvariationp+1formtheory}) $\delta \to\delta_{\chi}$,
integrating by parts and using the Noether identity $dE_{A}=0$, we find

\begin{subequations}
  \begin{align}
  \delta_{\chi} S[A]
   & =
     \int_{\mathcal{M}} d\mathbf{J}[\chi]\,,
    \\
    & \nonumber \\
    \mathbf{J}[\chi]
    & \equiv
      \mathbf{\Theta}(A,\delta_{\chi} A)
      +(-1)^{d-p-1}\mathbf{E}_{A}\wedge \chi\,.
  \end{align}
\end{subequations}

Since the action is exactly invariant under these transformations for any
choice of $\mathcal{M}$, we conclude that

\begin{equation}
  d\mathbf{J}[\chi]
  =
  0\,,
\end{equation}

\noindent
off-shell. This happens because

\begin{subequations}
  \begin{align}
    \mathbf{J}[\chi]
    & =
      d\mathbf{Q}[\chi]\,,
    \\
    & \nonumber \\
    \mathbf{Q}[\chi]
    & =
      (-1)^{d-p}\star F\wedge \chi\,.
  \end{align}
\end{subequations}

These results can be exploited to define charges that satisfy a Gauss law by
integrating the $(d-2)$-form $\mathbf{Q}[\chi]$ over compact
$(d-2)$-dimensional ``surfaces'' $\Sigma^{d-2}$

\begin{equation}
  q[\chi]
  =
  \int_{\Sigma^{d-2}}\mathbf{Q}[\chi]\,,
\end{equation}

\noindent
as follows \cite{Barnich:2003xg}: observe that, by construction

\begin{equation}
  d\mathbf{Q}[\chi]
  =
  \mathbf{J}[\chi]
  =
  \mathbf{\Theta}(A,\delta_{\chi} A)
      +(-1)^{d-p-1}\mathbf{E}_{A}\wedge \chi\,.
\end{equation}

The second term can be made to vanish by choosing a solution of the equations
of motion $\mathbf{E}_{A}=0$. Then, since the pre-symplectic potential is
linear in $\delta A$, the first term can be made to vanish by choosing a
\textit{Killing} or \textit{reducibility} parameter $\chi=\kappa$ such that

\begin{equation}
  \delta_{\kappa}A
  =
  d\kappa
  =
  0\,.
\end{equation}

Then, the charge $q[\kappa]$ satisfies a Gauss law because
$d\mathbf{Q}[\kappa]=0$. Again, for manifolds $\mathcal{M}$ with appropriate
topology we can write the most general closed $\kappa$ in the form

\begin{equation}
  \kappa
  =
  b^{I}h^{(p)}_{I}+de\,,
\end{equation}

\noindent
where the $b^{i}$ are constants and $\{ h^{(p)}_{I}\}$ is a basis of closed
but not exact $p$-forms. We can define an independent charge for each of them

\begin{equation}
  q_{I}
  \equiv
  q[h^{(p)}_{I}]
=
 (-1)^{d-p}\int_{\Sigma^{d-2}}\star F\wedge h^{(p)}_{I}\,,
\end{equation}

\noindent
but not for the exact part:

\begin{equation}
  \int_{\Sigma^{d-2}}\star F\wedge de
  \sim
  \int_{\Sigma^{d-2}}d\left(\star F\wedge e\right)
  =
  \int_{\partial\Sigma^{d-2}}\star F\wedge e
  =
  0\,,
\end{equation}

\noindent
where we have used the equations of motion and the compactness of the
integration surface. For this reason, the charges $q_{I}$ we have defined are
insensitive to the ambiguity of the $h^{(p)}_{I}$, which are defined only up
to the addition of a total derivative.

In the next section we are going to try to generalize this scheme in the
context of gravity.

\section{Higher-form symmetries in gravity}
\label{sec-highergrav}

The EH action is invariant up to total derivatives under the
following transformations of the metric field

\begin{equation}
  \delta_{\xi}g_{\mu\nu}
  =
  -\pounds_{\xi}g_{\mu\nu}
  =
  -2\nabla_{(\mu}\xi_{\nu)}\,,
\end{equation}

\noindent
which are obviously associated to diffeomorphisms
$\delta_{\xi} x^{\mu}=\xi^{\mu}$.

We would like to find more general transformations $\delta g_{\mu\nu}$ leaving
the action invariant, at least if certain geometrical or topological
conditions are met. In order to get some inspiration, let us consider the KK
theory with one compact and isometric direction which we discussed in the
introduction. In that case, both the transformations that become the gauge
transformations of the KK vector in one dimension less and the transformations
that rescale the KK vector and scalar in one dimension less are associated to
diffeomorphisms generated by vector fields of the form

\begin{equation}
  f(\hat{x})k\,,
  \hspace{1.5cm}
  k
  =
  \partial_{\underline{z}}\,.  
\end{equation}

\noindent
When the function $f(\hat{x})$ is an arbitrary function $\Lambda(x)$ of the
$(\hat{d}-1)$-dimensional coordinates $x$, only, we get the gauge
transformations of the KK vector field and when it is proportional to the
isomeric coordinate $z$, it generates rescalings of the $z$ coordinate. We
have argued that these rescalings are incompatible with the periodic boundary
conditions of $z$, but let us ignore this fact for the moment and let us look
at the transformations of the metric generated by vector fields of the above
form. They act on the metric as

\begin{equation}
  \delta_{f}\hat{g}_{\hat{\mu}\hat{\nu}}
  =
  -2\partial_{(\hat{\mu}}f k_{\hat{\nu})}\,,
\end{equation}

\noindent
because $k$ is a Killing vector that leaves invariant the metrics of the class
considered in KK theory.

When $f=z$, the above transformation is well defined, even though we derived
it from a vector field which is not. This fact suggests that we may try to
search for transformations of the metric $g_{\mu\nu}$\footnote{Now we consider
  a general metric in a general dimension and, henceforth, suppress the hats.}
of the form

\begin{equation}
  \delta_{\epsilon} g_{\mu\nu}
  \equiv
  -2\epsilon_{(\mu}k_{\nu)}\,,
\end{equation}

\noindent
where $\epsilon=\epsilon^{\mu}\partial_{\mu}$ is some vector field and
$k=k^{\mu}\partial_{\mu}$ is a Killing vector field of the metric, so that

\begin{equation}
  \label{eq:KVE}
  \nabla_{\mu}k_{\nu}
  =
  \nabla_{[\mu}k_{\nu]}\,.
\end{equation}

First of all, for the above transformation to be a symmetry in the class of
metrics admitting $k$ as Killing vector, the following condition must be
satisfied:

\begin{equation}
  \pounds_{k}\left(\epsilon_{(\mu}k_{\nu)}\right)
  =
  0\,.
\end{equation}

\noindent
This condition is equivalent to

\begin{equation}
  \pounds_{k}\hat{\epsilon}
  =
  \imath_{k}d\hat{\epsilon}+d\imath_{k}\hat{\epsilon}
  =
  0,
\end{equation}

\noindent
which would be satisfied if (but not only if)

\begin{subequations}
  \label{eq:consistencyconditions}
  \begin{align}
  d\hat{\epsilon}
  & =
  0\,,
    \\
    & \nonumber \\
    d\imath_{k}\hat{\epsilon}
    & =
      0\,.
  \end{align}
\end{subequations}

We now want to see under which conditions these transformations leave the EH
action invariant. We know that, when $\epsilon_{\mu}=\partial_{\mu}f$ or
$\hat{\epsilon}=df$\footnote{In this context we denote with hats the 1-form
  $\hat{\epsilon}=\epsilon_{\mu}dx^{\mu}$ dual to the vector field
  $\epsilon=\epsilon^{\mu}\partial_{\mu}$.}  for some well-defined $f$, the
above transformation is associated to a diffeomorphism, much in the same way
as the transformation $\delta A$ of a $(p+1)$-form field corresponds to a
gauge transformation when it is exact. Thus, we expect the transformation of
EH action to be proportional to $d\hat{\epsilon}$ and, perhaps, to vanish when
$\hat{\epsilon}$ is closed.

A straightforward calculation gives for the Levi-Civita connection

\begin{equation}
  \label{eq:deltaepsilonGamma}
  \begin{aligned}
    \delta_{\epsilon}\Gamma_{\mu\nu}{}^{\rho}
    & =
    -g^{\rho\sigma}
      \left\{
      k_{\sigma}\nabla_{(\mu}\epsilon_{\nu)}
      +k_{\nu}\nabla_{[\mu}\epsilon_{\sigma]}
      +k_{\mu}\nabla_{[\nu}\epsilon_{\sigma]}
      +2\epsilon_{(\mu}\nabla_{\nu)}k_{\sigma}
    \right\}\,,
  \end{aligned}
\end{equation}

\noindent
where we have used the Killing vector equation (KVE) Eq.~(\ref{eq:KVE}) in the
last step.

The variation of the Riemann curvature tensor is given by the Palatini
identity

\begin{equation}
  \begin{aligned}
    \delta_{\epsilon}R_{\alpha\mu\nu}{}^{\rho}
    & =
      2\nabla_{[\alpha|} \delta_{\epsilon}\Gamma_{|\mu]\nu}{}^{\rho}
    \\
    & \\
    & =
      -g^{\rho\sigma}
      \left\{
      \nabla_{[\alpha|}k_{\sigma}\nabla_{|\mu]}\epsilon_{\nu}
      +\nabla_{[\alpha|}k_{\sigma}\nabla_{\nu}\epsilon_{|\mu]}
      +\nabla_{[\alpha|}k_{\nu}\nabla_{|\mu]}\epsilon_{\sigma}
      -\nabla_{[\alpha|}k_{\nu}\nabla_{\sigma}\epsilon_{|\mu]}
      \right.
    \\
    & \\
    & \hspace{.5cm}
      \left.
      +2\nabla_{\alpha}k_{\mu}\nabla_{[\nu}\epsilon_{\sigma]}
      +2\nabla_{[\alpha}\epsilon_{\mu]}\nabla_{\nu}k_{\sigma}
      +2\nabla_{[\alpha|}\epsilon_{\nu}\nabla_{|\mu]}k_{\sigma}
      \right.
    \\
    & \\
    & \hspace{.5cm}
      \left.
      +k_{\sigma}\nabla_{[\alpha}\nabla_{\mu]}\epsilon_{\nu}
      +k_{\sigma}\nabla_{[\alpha|}\nabla_{\nu}\epsilon_{|\mu]}
      +k_{\nu}\nabla_{[\alpha}\nabla_{\mu]}\epsilon_{\sigma}
      -k_{\nu}\nabla_{[\alpha|}\nabla_{\sigma}\epsilon_{|\mu]}
      \right.
    \\
    & \\
    & \hspace{.5cm}
      \left.
      +2k_{[\mu}\nabla_{\alpha]}\nabla_{[\nu}\epsilon_{\sigma]}
      +2\epsilon_{[\mu}\nabla_{\alpha]}\nabla_{\nu}k_{\sigma}
      +2\epsilon_{\nu}\nabla_{[\alpha}\nabla_{\mu]}k_{\sigma}
      \right\}\,.
  \end{aligned}
\end{equation}

In order to simplify the notation we define

\begin{equation}
  P_{\epsilon\, \mu\nu}
  \equiv
  \nabla_{[\mu}\epsilon_{\nu]}\,,
  \hspace{1cm}
  P_{k\, \mu\nu}
  \equiv
  \nabla_{\mu}k_{\nu}\,.
\end{equation}

\noindent
Using the Ricci identity and the integrability condition of the Killing vector

\begin{subequations}
  \begin{align}
    [\nabla_{\mu},\nabla_{\nu}] \xi_{\rho}
    & =
    -R_{\mu\nu\rho}{}^{\sigma}\xi_{\sigma}\,,
    \\
    & \nonumber \\
    \nabla_{\mu}\nabla_{\nu}k_{\rho}
    & =
      -k^{\sigma}R_{\sigma\mu\nu\rho}\,,
  \end{align}
\end{subequations}

\noindent
we can remove all the second derivatives and the transformation of the Riemann
curvature tensor takes a much simpler form:

\begin{equation}
  \begin{aligned}
\delta_{\epsilon}R_{\alpha\mu\nu}{}^{\rho}
    & =
    -g^{\rho\sigma}
      \left\{
      -2P_{k\, [\alpha|\sigma}P_{\epsilon\, |\mu]\nu}
      +2P_{k\, [\alpha|\nu}P_{\epsilon\, |\mu]\sigma}
      +2P_{k\,\alpha\mu}P_{\epsilon\, \nu\sigma}
      +2P_{k\, \nu\sigma}P_{\epsilon\, \alpha\mu}
      \right.
    \\
    & \\
    & \hspace{.5cm}
      \left.
      -k_{\sigma}R_{\alpha\mu\nu\lambda}\epsilon^{\lambda}
      -2k_{\sigma}\nabla_{[\alpha|}P_{\epsilon\, |\mu]\nu}
      +2k_{\nu}\nabla_{[\alpha}P_{\epsilon\, |\mu]\sigma}
      \right.
    \\
    & \\
    & \hspace{.5cm}
      \left.
      +2k_{[\mu}\nabla_{\alpha]}P_{\epsilon\, \nu\sigma}
      -2\epsilon_{[\mu}k^{\lambda}R_{\lambda|\alpha]\nu\sigma}
      -2\epsilon_{\nu}k^{\lambda}R_{\lambda[\alpha\mu]\sigma}
    \right\}\,.
  \end{aligned}
\end{equation}

Replacing $\rho$ by $\mu$ and using

\begin{equation}
  \nabla_{[\mu|}P_{\epsilon\, |\nu\rho]}
  =
  0\,,
\end{equation}

\noindent
we get the transformation of the Ricci tensor

\begin{equation}
  \begin{aligned}
\delta_{\epsilon}R_{\alpha\nu}
    & =
      -6P_{k\,(\alpha|\mu}P_{\epsilon\, |\nu)}{}^{\mu}
      +2k_{(\alpha|}\nabla_{\mu}P_{\epsilon\, |\nu)}{}^{\mu}
      -2k_{\mu}\nabla_{(\alpha|}P_{\epsilon\, |\nu)}{}^{\mu}
      -2k^{\lambda}R_{\lambda(\alpha}\epsilon_{\nu)}\,.
  \end{aligned}
\end{equation}

Then

\begin{equation}
  \begin{aligned}
    \delta_{\epsilon}\left(\sqrt{|g|}\, R\right)
    & =
      \sqrt{|g|}\left[       -\delta_{\epsilon}g_{\mu\nu}G^{\mu\nu}
      +g^{\mu\nu}\delta_{\epsilon}R_{\mu\nu}\right]
    \\
    & \\
    & =
      -\epsilon^{\mu}k_{\mu}\, \sqrt{|g|}\, R
      +\sqrt{|g|}\left[
      -6\, P_{k}{}^{\mu\nu}P_{\epsilon\, \mu\nu}
      +4k_{\mu}\nabla_{\nu}P_{\epsilon}{}^{\mu\nu}
      \right]\,.
  \end{aligned}
\end{equation}

We can eliminate the second term by demanding

\begin{equation}
  \label{eq:epsilonclosed}
  P_{\epsilon\, \mu\nu}
  =
  0\,,
  \,\,\,\,\,
  \Rightarrow
  \,\,\,\,\,
  d\hat{\epsilon}
  =
  0\,,
\end{equation}

\noindent
which is the first of Eqs.~(\ref{eq:consistencyconditions}).  The first term
may be eliminated by demanding
$k^{\mu}\epsilon_{\mu}=\imath_{k}\hat{\epsilon} =0$, which is a very strong
condition. Instead, we observe that the result that we have obtained after
demanding the closedness of $\hat{\epsilon}$, Eq.~(\ref{eq:epsilonclosed}), is

\begin{equation}
  \delta_{\epsilon}\left(\sqrt{|g|}\, R\right)
  =
  -\imath_{k}\hat{\epsilon}\,\left( \sqrt{|g|}\, R\right)\,,
\end{equation}

\noindent
which would be equivalent to a global rescaling of the action if

\begin{equation}
  \imath_{k}\hat{\epsilon}
  =
  \text{constant}\,,
  \,\,\,\,\,\,
  \Rightarrow
  \,\,\,\,\,\,
  d\imath_{k}\hat{\epsilon}
  =
  0\,,
\end{equation}

\noindent
which is the second of the consistency conditions
Eqs.~(\ref{eq:consistencyconditions}).

We can compensate this rescaling with another global rescaling

\begin{equation}
  \delta_{\alpha}g_{\mu\nu}
  =
  \alpha g_{\mu\nu}\,,
  \,\,\,\,\,
  \Rightarrow
  \,\,\,\,\,
  \delta_{\alpha}\left(\sqrt{|g|}\, R\right)
  =
\frac{(d-2)}{2}\alpha\left(\sqrt{|g|}\, R\right)\,,
\end{equation}

\noindent
choosing the parameter $\alpha = 2 \imath_{k}\hat{\epsilon}/(d-2)$. Thus,
combining these two transformations, we find that

\begin{equation}
  \label{eq:highersymmetriesEH}
  \delta_{\epsilon}g_{\mu\nu}
  =
  -2 \epsilon_{(\mu}k_{\nu)}+ \frac{2}{(d-2)}\epsilon^{\rho}k_{\rho}g_{\mu\nu}\,,
\end{equation}

\noindent
leaves the EH action exactly invariant if the consistency
conditions Eqs.~(\ref{eq:consistencyconditions}) are met.

In the next section we are going to see in the KK context that these
transformations include and extend the diffeomorphism invariance of the
EH action and that, when they are not diffeomorphisms, they
generate the constant rescalings of the KK vector and scalar that leave
invariant the $d=(\hat{d}-1)$-dimensional action as well as the
$\hat{d}$-dimensional one, as we have shown. Notice that, in the KK setting the
dimensional parameter that occurs in the above formulae has to be replaced by
$\hat{d}$.

\subsection{Conserved charges}
\label{sec-conservedcharges}

We have just shown that, under the conditions
Eqs.~(\ref{eq:consistencyconditions}), the transformations
Eq.~(\ref{eq:highersymmetriesEH}) leave the EH action exactly invariant

\begin{equation}
  \delta_{\epsilon}S_{EH}[g]
  =
  0\,.
\end{equation}

\noindent
Associated to each of the closed, but not exact, 1-forms $\hat{\epsilon}$
satisfying also $d\imath_{k}\epsilon=0$, $\{\epsilon_{i}\}$, there must be a
conserved Noether current.

Under a general variation of the metric, 

\begin{subequations}
  \begin{align}
    \delta S_{EH}[g]
    & \sim
      \int_{\mathcal{M}}d^{4}x\,\left\{\frac{\delta S}{\delta g_{\mu\nu}}\delta
      g_{\mu\nu} +\partial_{\mu}\Theta^{\mu}(g,\delta g)\right\}\,,
    \\
    & \nonumber \\
    \frac{\delta S}{\delta g_{\mu\nu}}
    & =
      -\sqrt{|g|} G^{\mu\nu}\,,
    \\
    & \nonumber \\
    \Theta^{\mu}(g,\delta g)
    & =
      \sqrt{|g|}\left[g^{\mu\nu}\delta \Gamma_{\rho\nu}{}^{\rho}
      -g^{\rho\nu}\delta \Gamma_{\rho\nu}{}^{\mu}\right]\,.
  \end{align}
\end{subequations}

Then, for each $\epsilon_{i}$,

\begin{equation}
  \partial_{\mu}\Theta^{\mu}(g,\delta_{\epsilon_{i}} g)
  =
  -\frac{\delta S}{\delta g_{\mu\nu}}\delta_{\epsilon_{i}} g_{\mu\nu}
  \doteq
  0\,.
\end{equation}

\noindent
Then, using Eq.~(\ref{eq:deltaepsilonGamma}) and the fact that global
rescalings of the metric leave invariant the connection, the current
$j_{i}{}^{\mu}$ is given by

\begin{equation}
  \label{eq:conservedcurrent}
  \begin{aligned}
    j_{i}{}^{\mu}
    & =
      \Theta^{\mu}(g,\delta_{\epsilon_{i}} g)
    \\
    & \\
    & =
      \sqrt{|g|}\left\{
      k^{\mu}\nabla_{\rho}\epsilon_{i}{}^{\rho}
      +2\epsilon_{i}{}^{\rho}\nabla_{\rho}k^{\mu}
      \right\}\,,  
  \end{aligned}
\end{equation}

\noindent
where we have used the Killing equation satisfied by $k$ and the consistency
conditions Eqs.~(\ref{eq:consistencyconditions}).\footnote{For instance:
\begin{equation}
  -k^{\rho}\nabla_{\rho}\epsilon_{i}{}^{\mu}
  =
  -k^{\rho}\nabla^{\mu}\epsilon_{i\, \rho}
  =
  -\nabla^{\mu}\left(k^{\rho}\epsilon_{i\, \rho}\right)
  +\epsilon_{i}{}^{\rho}\nabla^{\mu}k_{\rho}
  =
  0-\epsilon_{i}{}^{\rho}\nabla_{\rho}k^{\mu}
  =
  -\epsilon_{i}{}^{\rho}\nabla_{\rho}k^{\mu}
  \,.
\end{equation}
}

We can check that, indeed, the above currents are conserved on-shell:

\begin{equation}
  \begin{aligned}
    \nabla_{\mu}\left(j_{i}{}^{\mu}/\sqrt{|g|}\right)
    & =
      -2\epsilon_{i}{}^{\rho}k^{\lambda}R_{\lambda\rho}
    \\
    & \\
    & \doteq
      0\,,
  \end{aligned}
\end{equation}

\noindent
where we have used

\begin{equation}
    \begin{aligned}
      k^{\mu}\nabla_{\mu}\nabla_{\rho}\epsilon_{i}{}^{\rho}
      & =
        0\,.
    \end{aligned}
  \end{equation}

  \section{Higher-form symmetries in the Kaluza--Klein setting}
  \label{sec-higherKK}

  Let us now consider the transformations that we have found in the preceding
  section in the KK setting in which one of the dimensions is compactified in
  a circle, parametrized by $z\in [0,2\pi \ell]$, which is the coordinate
  adapted to an isometry, so there is always a Killing vector field
  $k=\partial_{\underline{z}}$. In this setting, the solution to the first of
  the consistency conditions Eqs.~(\ref{eq:consistencyconditions}) is of the
  form\footnote{Observe that despite its local form, $dz$ is not an exact
    1-form because $z$ is not a single-valued function. Strictly speaking, it
    is not a good coordinate, either: S$^{1}$ needs to be covered by, at
    least, two coordinate patches with coordinates $z_{1}$ and $z_{2}$ related
    by an additive constant in the overlap. In each patch, the 1-form we are
    denoting by $dz$ would be $dz_{1}$ and $dz_{2}$ and it would be closed,
    but there is not a single-valued function $f$ such that it is $df$
    (exact). Most of the time it is simpler and sufficient to work with a
    single, periodically-identified, coordinate $z$ if one is careful.}

\begin{equation}
  \hat{\epsilon}
  =
  \beta dz +d\Lambda,
\end{equation}

\noindent
where $\beta$ is some constant. Then, the second equation reads

\begin{equation}
  \beta +\partial_{\underline{z}}\Lambda
  =
  \alpha\,,
\end{equation}

\noindent
and is solved by

\begin{equation}
  \Lambda
  =
  \Lambda(x)+(\alpha-\beta)z\,,
\end{equation}

\noindent
where $\partial_{\underline{z}}\Lambda(x)=0$. The second term in $\Lambda$
actually gives a term of the same kind as the first. Furthermore, as we have
discussed in the introduction, this second term is not single valued around
the circle and, therefore, we must remove it setting $\alpha = \beta$. Thus,

\begin{equation}
  \hat{\epsilon}
  =
  \alpha dz +d\Lambda(x)\,.
\end{equation}

The exact part of $\hat{\epsilon}$ is associated to a well-defined vector
field

\begin{equation}
  \lambda
  \equiv
  \Lambda(x)\partial_{\underline{z}}\,,  
\end{equation}

\noindent
which generates the gauge transformations of the KK vector field:





\begin{subequations}
  \begin{align}
    \delta_{\lambda}g_{\mu \nu}
    & =
      \delta_{\lambda}\left(\hat{g}_{\mu \nu}
      -\hat{g}_{\underline{z}\mu}\hat{g}_{\underline{z}\nu}/\hat{g}_{\underline{z}\underline{z}}\right)
      =
      0\,,
    \\[4pt]
    \delta_{\lambda}A_{\mu}
    & =
      \delta_{\lambda}\left(\hat{g}_{\mu \underline{z}}/\hat{g}_{\underline{z}\underline{z}}\right)
      =
      -\partial_{\mu}\Lambda\,,
    \\[4pt]
    \delta_{\lambda}k
    & =
    \delta_{\lambda}|\hat{g}_{\underline{z} \underline{z}}|^{1/2}
    =
      0\,.
  \end{align}
\end{subequations}

The global part of $\hat{\epsilon}$ acts on the $\hat{d}$-dimensional metric
as

\begin{equation}
  \delta_{\alpha}\hat{g}_{\hat{\mu}\hat{\nu}}
    =
      -2 \alpha \delta^{\underline{z}}{}_{(\hat{\mu}}
      \hat{g}_{\hat{\nu})\underline{z}}
      +\frac{2}{(\hat{d}-2)}\alpha \hat{g}_{\hat{\mu}\hat{\nu}}\,.
\end{equation}





These transformations act in a non-trivial way over all the fields in the
$d$-dimensional KK frame. In the Einstein frame, though, only the KK scalar
and vector transform

\begin{equation}
  \begin{aligned}
    \delta_{\alpha}g_{E\, \mu \nu}
    & =
\delta_{\alpha}\left(k^{2/(\hat{d}-3)}g_{\mu\nu}\right)
=
0\,,
    \\[4pt]
     \delta_{\alpha}A_{\mu}
    & =
      \alpha A_{\mu}\,,
    \\[4pt]   
    \delta_{\alpha}k
    & =
      -\alpha\frac{(\hat{d}-3)}{(\hat{d}-2)}k\,,
  \end{aligned}
\end{equation}

\noindent
and these transformations coincide precisely with those in
Eq.~(\ref{eq:alphatransformationsAk}) that leave the Einstein-frame action
Eq.~(\ref{eq:Einsteinframeactioncirclecompactification}) invariant.

Thus, we have shown that this global symmetry of the compactified theory is
related to a symmetry of the higher-dimensional EH action which is not a
diffeomorphism. This symmetry has been constructed with the help of a global
rescaling of the metric, but we are going to see that in more general toroidal
compactifications we do not need this global rescaling and we have symmetries
originating only in the global part of $\hat{\epsilon}$.

\subsection{Toroidal compactifications}

In toroidal compactifications there are $n$ compact, mutually commuting
isometries generated by Killing vectors that can be expressed in adapted
coordinates as 

\begin{equation}
  k_{m}
  =
  k_{m}{}^{\hat{\mu}}\partial_{\hat{\mu}}
  =
  \partial_{\underline{z}^{m}}
  \equiv
  \partial_{m}\,.
\end{equation}

The isometric coordinates $z^{m}$ parametrize the $n$ circles and, for
simplicity, we assume that all of them have the same period
$z^{m}\sim z^{m}+2\pi \ell$. The dimensional reduction of the
$\hat{d}$-dimensional EH action can be performed in two steps. First, we
decompose the $\hat{d}$-dimensional metric $\hat{g}_{\hat{\mu}\hat{\nu}}$ into
$d=(\hat{d}-n)$-dimensional fields: a KK-frame metric $g_{\mu\nu}$, $n$ gauge
fields (\textit{KK vectors}) $A^{m}=A^{m}{}_{\mu}dx^{\mu}$ and $n(n+1)/2$
\textit{KK scalar fields} described by a symmetric, positive-definite matrix
$G_{mn}$

\begin{equation}
   d\hat{s}^{2}
  =
  \hat{g}_{\hat{\mu}\hat{\nu}}d\hat{x}^{\hat{\mu}}d\hat{x}^{\hat{\nu}}
  =
 ds^{2}-G_{mn}(dz^{m}+A^{m})(dz^{n}+A^{n})\,,
\end{equation}

\noindent
where

\begin{equation}
  ds^{2}
  =
  g_{\mu\nu}dx^{\mu}dx^{\nu}\,.
\end{equation}

After integration over the $n$ isometric directions, we obtain the
$d$-dimensional KK-frame action

\begin{equation}
  S[g,A^{m},G_{mn}]
  \sim 
\int d^{d}x \sqrt{|g|}\, K
\left\{ R(g)
  -(\partial\ln{K})^{2}
-\textstyle\frac{1}{4} \partial_{\mu} G_{mn}\partial^{\mu}G^{mn}
-\textstyle\frac{1}{4} F^{2}
\right\}\,,
\end{equation}

\noindent
where

\begin{equation}
\label{eq:toroidal2}
K^{2}
\equiv
|\mathrm{det}(G_{mn})|\,,
\hspace{1cm}
F^{2}
\equiv
G_{mn}F^{m\, \mu\nu }F^{n}{}_{\mu\nu}\,,
\hspace{1cm}
F^{m}{}_{\mu\nu}
\equiv
2\partial_{[\mu}A^{m}{}_{\nu]}\,.
\end{equation}

The second step consists in a rescaling of the KK-frame metric to the
Einstein-frame metric. It is also convenient to rescale the matrix of scalars
to obtain a unimodular matrix $\mathcal{M}$:

\begin{equation}
  \begin{aligned}
    g_{\mu\nu}
    & =
      K^{-\frac{2}{d-2}}  g_{E\, \mu\nu}\,,
    \\
    & \\
    G_{mn}
    & \equiv
    K^{\frac{2}{n}}\mathcal{M}_{mn}\,.
  \end{aligned}
\end{equation}

The result is the following $d$-dimensional Einstein-frame action

\begin{equation}
\label{eq:torusKKaction}
\begin{aligned}
  S[g,A^{m},K,\mathcal{M}_{mn}]
  & \sim
      \int d^{d}x \sqrt{|g_{E}|}\, 
\left\{ R_{E} 
+\tfrac{1}{2}\left(\partial\ln{K^{-2a}}\right)^{2}
-\tfrac{1}{4} \partial_{\mu} \mathcal{M}_{mn}\partial^{\mu} 
    \mathcal{M}^{mn}
    \right.
  \\
&  \\
& 
\hspace{3.5cm}
\left.
-\tfrac{1}{4} K^{(2a)^{2}}
\mathcal{M}_{mn}F^{m\, \mu\nu}F^{n}{}_{\mu\nu} 
        \right\}\,,
\end{aligned}
\end{equation}

\noindent
where we have defined the  constant

\begin{equation}
  a
  \equiv
  -\sqrt{\frac{(d-2+n)}{2n(d-2)}}\,.
\end{equation}

This action is invariant under SL$(n,\mathbb{R})$ transformations which only
act on the KK vectors and $\mathcal{M}$

\begin{equation}
  A^{\prime\, m}
  =
  S^{m}{}_{n}A^{n}\,,
  \hspace{1cm}
  \mathcal{M}'_{mn}
  =
  S^{-1\, p}{}_{m}S^{-1\, q}{}_{n}
  \mathcal{M}_{pq}\,,
\end{equation}

\noindent
and which are only non-trivial for $n\geq 2$ and under global rescalings which
only act on the KK scalar $K$ and the KK vectors $A^{m}$

\begin{equation}
  \label{eq:Ctransformations}
  K'
  =
  C^{-\frac{1}{2a^{2}}}K\,,
  \hspace{1cm}
  A^{\prime\, m}
  =
  C A^{m}\,,
\end{equation}

\noindent
where $C$ is an arbitrary positive real constant. Together, they generate the
GL$(n,\mathbb{R})$ duality group of this theory which, in the literature, is
customarily associated to diffeomorphisms of the internal space
T$^{n}$. However, just as in the $n=1$ case, most of those transformations do
not preserve the boundary conditions and are generated by vector fields which
are not single valued.

In this case, we can consider a linear combination of transformations of the
form Eq.~(\ref{eq:highersymmetriesEH}) for each of the $n$ Killing vectors. We
have to introduce $n$ closed 1-forms $\hat{\epsilon}^{m}$ in order to
construct transformations
$\delta_{\epsilon}g_{\mu\nu}\sim \epsilon^{(m)}{}_{(\mu|}k_{(m)\, |\nu)}$, (no
sum over $m$ intended) but, once we have introduced them, we can obviously
consider other pairings $\epsilon^{m}{}_{(\mu|}k_{n\, |\nu)}$, $m\neq n$ and,
therefore, we are led to consider the most general possibility

\begin{equation}
  \label{eq:mostgeneral}
  \delta_{\epsilon}\hat{g}_{\hat{\mu}\hat{\nu}}
  =
  -2 T^{m}{}_{n}\epsilon^{n}{}_{(\hat{\mu}|}k_{m\, |\hat{\nu})}\,,  
\end{equation}

\noindent
where $T^{m}{}_{n}$ is a matrix of constant, infinitesimal, parameters. Each
of the terms in the linear combination of the right-hand side must satisfy the
condition

\begin{equation}
  \epsilon^{n\, \hat{\rho}}k_{m\, \hat{\rho}}
  =
  \text{constant}\,,
\end{equation}

\noindent
and, furthermore, it needs to be compensated by a global rescaling in order to
generate a symmetry of the EH action. Thus, we must consider the
transformations Eq.~(\ref{eq:mostgeneral}) supplemented by a global rescaling,
if necessary.

In this setting, we can always choose the closed 1-forms $\epsilon^{m}$ so
that

\begin{equation}
  \epsilon^{n\, \hat{\rho}}k_{m\, \hat{\rho}}
  =
  \delta^{n}{}_{m}\,,
  \,\,\,\,\,\,
  \Rightarrow
  \,\,\,\,\,\,
  \hat{\epsilon}^{m}
  =
  dz^{m}+d\Lambda^{m}(x)\,.
\end{equation}

\noindent
The exact part does not need to be supplemented by global rescalings. It
generates gauge transformations of the KK vectors

\begin{equation}
  \delta_{\Lambda}A^{m}
  =
  d\Lambda^{\prime\, m}\,,
  \hspace{1cm}
  \Lambda^{\prime\, m}
  =
  T^{m}{}_{n}\Lambda^{n}\,.
\end{equation}

In what follows, we will only consider the non-exact part.  Taking into
account the necessary global rescalings, the transformations take the form

\begin{equation}
  \delta_{\epsilon}\hat{g}_{\hat{\mu}\hat{\nu}}
  =
  -2 T^{p}{}_{q}\epsilon^{q}{}_{(\hat{\mu}|}k_{p\, |\hat{\nu})}
  +\frac{2}{(\hat{d}-2)}T^{p}{}_{p}\hat{g}_{\hat{\mu}\hat{\nu}}\,.
\end{equation}

Since only the trace of $T^{m}{}_{n}$ needs to be compensated by the global
rescalings, we decompose it into its traceless and trace parts:

\begin{equation}
  T^{m}{}_{n}
  =
  T^{m}{}_{n}-\frac{1}{n}\delta^{m}{}_{n}T^{p}{}_{p}
  +\frac{1}{n}\delta^{m}{}_{n}T^{p}{}_{p}
  \equiv
  R^{m}{}_{n}+T\delta^{m}{}_{n}\,,
\end{equation}

\noindent
and we end up with

\begin{equation}
  \delta_{\epsilon}\hat{g}_{\hat{\mu}\hat{\nu}}
  =
  -2 R^{p}{}_{q}\epsilon^{q}{}_{(\hat{\mu}|}k_{p\, |\hat{\nu})}
-2T\left[\epsilon^{p}{}_{(\hat{\mu}|}k_{p\, |\hat{\nu})}
  -\frac{n}{(\hat{d}-2)}\hat{g}_{\hat{\mu}\hat{\nu}}\right]\,.
\end{equation}

We can define two independent sets of transformations:

\begin{subequations}
  \begin{align}
    \delta_{R}\hat{g}_{\hat{\mu}\hat{\nu}}
    & \equiv
      -2 R^{p}{}_{q}\epsilon^{q}{}_{(\hat{\mu}|}k_{p\, |\hat{\nu})}\,,
      \hspace{1cm}
      R^{p}{}_{p}=0\,,
    \\
    & \nonumber \\
    \delta_{T}\hat{g}_{\hat{\mu}\hat{\nu}}
    & \equiv
    -2T\left[\epsilon^{p}{}_{(\hat{\mu}|}k_{p\, |\hat{\nu})}
  -\frac{n}{(\hat{d}-2)}\hat{g}_{\hat{\mu}\hat{\nu}}\right]\,.
  \end{align}
\end{subequations}

Taking into account that

\begin{equation}
  \epsilon^{p}{}_{(\hat{\mu}|}k_{q\, |\hat{\nu})}
    =
      \delta^{p}{}_{(\hat{\mu}|}\hat{g}_{q\, |\hat{\nu})}\,,
\end{equation}
      
\noindent
the effect of the $\delta_{T}$ transformations
%
%
on the $d$-dimensional Einstein-frame fields is

\begin{equation}
  \begin{aligned}
    \delta_{T}K
    & =
      -\frac{n(\hat{d}-2-n)T}{(\hat{d}-2)} K
      =
      -\frac{n(d-2)T}{(d-2+n)} K
      =
      -\frac{T}{2a^{2}}K\,,
    \\[4pt]
    \delta_{T}A^{m}{}_{\mu}
    & =
      T A^{m}{}_{\mu}\,,
    \\[4pt]
  \end{aligned}
\end{equation}

\noindent
and the Einstein metric and $\mathcal{M}_{mn}$ are invariant. These
transformations are the infinitesimal version of those in
Eq.~(\ref{eq:Ctransformations}), which leave the Einstein-frame action
invariant.

The $\delta_{R}$ transformations act on the $d$-dimensional fields as


\begin{equation}
  \begin{aligned}
    \delta_{R}\mathcal{M}_{mn}
    & =
      -2 R^{p}{}_{(m|}\mathcal{M}_{p\, |n)}
      \,,
      \\[4pt]
    \delta_{R}A^{m}{}_{\mu}
    & =
      R^{m}{}_{n}A^{n}{}_{\mu}\,,
  \end{aligned}
\end{equation}

\noindent
leaving the rest invariant. 

\subsection{Conserved charges}
\label{sec-conservedchargesKK}

Let us compute the $\hat{d}$-dimensional Noether current of the $T$ and $R$
symmetries. Using the general expression Eq.~(\ref{eq:conservedcurrent}), we
find

\begin{equation}
  \begin{aligned}
    j^{m}{}_{n}{}^{\hat{\mu}}/\sqrt{|\hat{g}|}
    & =
      k_{n}{}^{\hat{\mu}}\hat{\nabla}_{\hat{\rho}}\epsilon^{m\, \hat{\rho}}
      +2\epsilon^{m\,\hat{\rho}}\hat{\nabla}_{\hat{\rho}}k_{n}{}^{\hat{\mu}}
    \\
    & \\
    & =
      -\delta_{n}{}^{\hat{\mu}}\frac{1}{\sqrt{|g|}\,K}\partial_{\rho}\left(\sqrt{|g|}\,  K\, A^{m\, \rho}\right)
      -2 A^{m\,\rho}\hat{\Gamma}_{\rho n}{}^{\hat{\mu}}
    \\
    & \\
    & \hspace{.5cm}
      -2 \left(G^{mp}-A^{m}{}_{\rho}A^{p\, \rho}\right)
      \hat{\Gamma}_{pn}{}^{\hat{\mu}}\,,
  \end{aligned}
\end{equation}

\noindent
where we have decomposed the $\hat{d}$-dimensional fields in terms of the
$d$-dimensional ones. The $d$-dimensional components are

\begin{equation}
  \begin{aligned}
    j^{m}{}_{n}{}^{\mu}/\sqrt{|\hat{g}|}
    & =
      K^{\frac{2}{d-2}}\left[
      K^{(2a)^{2}}\mathcal{M}_{np}F^{p}{}_{\rho}{}^{\mu}A^{m\,\rho}
      -\mathcal{M}^{mp}\partial^{\mu}\mathcal{M}_{pn}
      -\frac{2}{n}K^{-1}\partial^{\mu}K \delta^{m}{}_{n}\right]\,.
  \end{aligned}
\end{equation}

\noindent
where we have expressed the current in the Einstein-frame metric.

The trace part is, (again in the Einstein frame)

\begin{equation}
    j^{m}{}_{m}{}^{\mu}/\sqrt{|g|}
    \sim
      2K^{-1}\partial^{\mu}K
      +K^{(2a)^{2}}\mathcal{M}_{mn}F^{m\, \mu}{}_{\rho}A^{n\,\rho}\,,
\end{equation}

\noindent
and it is easy to check that its divergence is proportional a combination of
the $d$-dimensional equations of motion of $K$ and $A^{m}$ and vanishes
on-shell.

The traceless part is

\begin{equation}
R^{n}{}_{m} j^{m}{}_{n}{}^{\mu}/\sqrt{|g|}
\sim
R^{n}{}_{m}\left[
  \mathcal{M}^{mp}\partial^{\mu}\mathcal{M}_{pn}
  +K^{(2a)^{2}}\mathcal{M}_{np}F^{p\, \mu}{}_{\rho}A^{m\,\rho}
  \right]\,,
\end{equation}

\noindent
and its divergence is a combination of the equations of motion of $A^{m}$
and

\begin{equation}
R^{p}{}_{m}\mathcal{M}_{np}\frac{\delta S}{\delta \mathcal{M}_{mn}}\,,  
\end{equation}

\noindent
which, therefore, also vanishes on-shell.

Both currents coincide with the Noether currents of the $T$ and $R$ symmetries
of the $d$-dimensional action, as expected.

\section{Discussion}
\label{sec-discussion}

In this paper we have shown that the Einstein--Hilbert action is invariant
under transformations of the metric which are not diffeomorphisms. In the KK
stting that we have chosen as an example, these transformations are equivalent
to diffeomorphisms which are not globally well defined,\footnote{This is
  always going to be the case, since all closed 1-forms are locally exact.}
and they give well-known symmetries of the compactified theory. Nevertheless,
these symmetries provide a higher-dimensional explanation for them which,
strictly speaking, was not available in the literature. On the other hand, we
believe that the symmetries that we have found are just the simplest in their
class and that, in more general settings, the Einstein--Hilbert action will
certainly admit more general non-diffeomorphic symmetries.

There is another interesting aspect of the relation between global duality
symmetries in compactified theories and non-diffeomorphic symmetries in higher
dimensions. It is believed that Quantum Gravity theories should not have any
global symmetries
\cite{Banks:1988yz,Banks:2010zn,Harlow:2018jwu,Harlow:2018tng}, but if these
symmetries were just a global subgroup of a gauge group, it would be very
difficult to argue that only that particular subset should be broken. Our
results imply (at least in the simple examples tat we have explored here) that
those global duality symmetries are not a subgroup of the group of
diffeomorphisms and, therefore, they can be broken while preseving the
integrity of the group of diffeomorphisms.

It should also be clear that the coupling to matter may modify or enhance the
set of non-diffeomorphic symmetries, mixing now different higher-dimensional
fields as it happens in T~duality.

Finally, we know that, when we compactify a theory, the existence of global
symmetries in a theory allows for \textit{generalized dimensional reduction}
ansatzs in which one performs a global symmetry transformation of the matter
fields with a parameter which is linear in one of the compact
coordinates. These ansatzs lead to gauge/massive theories in lower
dimensions. A good example is provided by the generalized dimensional
reduction of type~IIB supergravity from 10 to 9 dimensions exploting the full
SL$(2,\mathbb{R})$ global symmetry of the theory performed in
Ref.~\cite{Meessen:1998qm}. These generalized dimensional reductions can also
be associated to the introduction of non-dynamical branes in the background
\cite{Bergshoeff:1996ui,Meessen:1998qm}. There is another kind of generalized
(\textit{Scherk--Schwarz}) dimensional reduction ansatz, proposed in
Ref.~\cite{Scherk:1979zr}, which may be related to the kind of global
symmetries acting on the metric that we have been discussing in this paper. In
future work we would like to explore the possible connection between the
Scherk--Schwarz ansatz and the global, non-diffeomorphic symmetries identified
in this paper.

\section*{Acknowledgments}

TO would like to thank J.J.~Fern\'andez-Melgarejo and A.~Rosabal for
interesting conversations and for their hospitality at the U.~of Murcia.  This
work has been supported in part by the MCIU, AEI, FEDER (UE) grant
PID2021-125700NB-C21 (``Gravity, Supergravity and Superstrings'' (GRASS)) and
IFT Centro de Excelencia Severo Ochoa CEX2020-001007-S. The work of CG-F was
supported by the MU grant FPU21/02222. The work of MZ was supported by the
fellowship LCF/BQ/DI20/11780035 from ``la Caixa'' Foundation (ID
100010434). TO wishes to thank M.M.~Fern\'andez for her permanent support.

\appendix


\end{document}